==================
\documentstyle [12pt]{article}
\title{
\hspace{4.0truein}{\small UBCTP-92-24}\\
\hspace{4.0truein}{\small UCSBTH-92-24}\\
\vspace{0.30truein}
{Driving Operators Relevant: \\A Feature of Chern-Simons Interaction}}
\author{Wei Chen\footnotemark[1]  \\
  Department of Physics, University of British Columbia\\
          Vancouver, B.C. Canada V6T 1Z1\\
              Miao Li\footnotemark[2]\\
       Department of Physics, University of California\\
          Santa Barbara, CA 93106 USA }
\footnotetext{$^*$ chen@physics.ubc.ca}
\footnotetext{$^\dagger$ Present address: Department of
Physics, Brown University, Providence, RI 02912 USA. li@het.brown.edu}
\vspace{0.2truein}
\date{September 1992}

\begin{document}
\maketitle

\vspace{0.1truein}

\begin{abstract}

By computing anomalous dimensions
of gauge invariant composite operators $(\bar\psi\psi)^n$ and
$(\phi^*\phi)^n$ in Chern-Simons fermion and boson models,
we address that Chern-Simons interactions make these operators more
relevant or less irrelevant in the low energy region.
We obtain a critical
Chern-Simons fermion coupling, ${1\over \kappa_c^2} = {6\over 19}$,
for a phase transition at which the leading irrelevant four-fermion
operator $(\bar\psi\psi)^2$ becomes marginal,
and a critical Chern-Simons boson coupling,
${1\over \kappa_c^2} = {6\over 34}$, for a similar phase transition for
the leading irrelevant operator $(\phi^*\phi)^4$.
We see this phenomenon also in the $1/N$ expansion.

\end{abstract}

\newpage
\baselineskip=20.truept

One of the remarkable features of Chern-Simons matter gauge
theories \cite{DJT}\ is that the Chern-Simons interaction attaches
statistical flux tubes to particles, by which
a fermion can be transmuted into a boson or an anyon (and vise versa),
depending on the strength of the Chern-Simons coupling,
$1/\kappa$ ($\kappa$ the statistical parameter) \cite{P}.
In the relativistic quantum field theory scheme,
an Abelian Chern-Simons term
receives {\it no} correction
from interacting with massive matters beyond one loop\cite{CH}\cite{SSW},
and receives only {\it finite} corrections from massless matters
starting from two loops \cite{SSW}\cite{C}.
This results in an identically vanishing beta function
for Chern-Simons couplings, as the theories are gauge invariant.
The insensibility of Chern-Simons couplings
to energy scales is attributed to the topological nature
of Chern-Simons actions. However, it
doesn't imply a triviality of the whole
theory in the sense of renormalization.
In fact, as shown in \cite{CSW},
matter fields in Chern-Simons models {\it do} need infinite
renormalization and receive
anomalous dimensions%
\footnotemark.
\footnotetext{Non-Abelian Chern-Simons field receives
an anomalous dimension through interacting
with matters as well \cite{CSW}, while the Chern-Simons coefficient
keeps finite and quantized, as required by the ``large'' gauge symmetry.}
This observation leads naturally to a conjecture
that asymptotic behavior of gauge invariant operators in a
Chern-Simons matter
theory is non-trivial.
Recently, Chern-Simons boson and fermion models have been used
in describing the phase transition between quantum Hall states
and insulators in \cite{WW} and \cite{FCW}, respectively,
where among other observations is that
the composite mass operators of matter fields
receive as well an anomalous
dimension, and therefore not only the critical
exponent $\eta$ but also $\nu$ is modified
by the Chern-Simons interaction.

In the present letter, we shall address another feature of Chern-Simons
matter interaction, namely it modifies the
scaling dimensions of a class of gauge invariant composite operators
$(\bar\psi\psi)^n$ of fermion model and
$(\phi^*\phi)^n$ of boson model, with $n = positive~integers$,
in such a way that these operators are driven more relevant or
less irrelevant (in the lower energy region).

By dimensional analysis, not many operators
in a Chern-Simons Matter model are relevant.
For instance, the Euclidean action of
a Chern-Simons fermion model, if contains the relevant and marginal operators
only, reads
\begin{equation}
S = \int d^3x \left( \bar\psi\gamma\cdot\partial\psi
 + m\bar\psi\psi + \bar\psi\gamma\cdot A\psi
 + i{\kappa \over 4\pi}\epsilon^{ijk}A_i\partial_j A_k\right).
\label{action}
\end{equation}
The action (\ref{action}) is so normalized that $\kappa$ is dimensionless.
In particular, the four-fermion operator
$(\bar\psi\psi)^2$ is irrelevant as it has an engineering dimension {\it four};
and  the Maxwell term $F^2$ ($F$ the field strength),
the kinetic of a dynamic gauge field, is irrelevant,
compared to the kinetic Chern-Simons $AdA$.
Now, an interesting question is
how will these be changed by quantum
corrections? The answer to the operator $F^2$ is obvious:
it remains irrelevant, as the dimension of the primary gauge
field $A$ doesn't change during the quantization.
However, the radiation correction from matters induces
a non-local operator concerning the gauge field $A$
 in the effective Lagrangian. In the case of
massless matters for instance, this induced term is
$F(1/\sqrt{\partial^2})F$.
It is finite and marginal.
On the other hand, the operator $(\bar\psi\psi)^2$ and
other gauge invariant composite operators $(\bar\psi\psi)^n$
are expected to have an anomalous dimension. One of the main
results of the present work is the scaling
dimensions of such a class of composite operators
at the lowest non-trivial order
\begin{equation}
d_{(\bar\psi\psi)^n} = 2n - {2n+15\over 6\kappa^2}
+ O({1\over \kappa^4}).
\label{dn}
\end{equation}
(\ref{dn}) shows that the Chern-Simons interaction
makes the mass operator ($n=1$), already relevant
without the Chern-Simons interaction,
even more relevant.
Its consequences in some phase transitions have been discussed
in \cite{FCW}.
Moreover, also due to the Chern-Simons interaction,
the irrelevant operators $(\bar\psi\psi)^n$ ($n\geq 2$) are
less irrelevant.
The most important operator of this type is
the leading irrelevant one, with $n=2$.
(\ref{dn})\ suggests a phase transition with
a critical effective coupling
\begin{equation}
1/\kappa_c^2 = 6/19,
\label{kc}
\end{equation}
at which the leading irrelevant four-fermion operator $(\bar\psi\psi)^2$
becomes marginal. The critical value in (\ref{kc}) shows that if such a phase
transition would occur, it
occurs only in a region with rather strong Chern-Simons
interaction. We make several remarks at this point.
First, it is not difficult to check that, at the order $1/k^3$, all
associated Feynman diagrams are {\it finite}, the next leading
contributions to the anomalous
dimensions are from the order $1/\kappa^4$ (four and higher loops),
as indicated in (\ref{dn}). According to (\ref{kc}), therefore,
there exists a multiplicative factor $1/\kappa_c^4\simeq 1/10$
in the next order.
If the numerical coefficient of the next order
would be smaller than or comparable to that of the
leading order, the perturbation expansion near the transition point
is, to some extent, acceptable.  Our second remark is that,
once such a phase transition happens,
a four-fermion operator, now marginal or relevant,
is switched on in the system. In turn, this four-fermion
self-interaction develops a {\it positive} anomalous dimension
\cite{CMS}~that has the potential to make the operator irrelevant.
As a result of
competition between the Chern-Simon-fermion and
four-fermion interactions, a balance would be reached
somewhere. The last, a bare four-fermion interaction in three
dimensions has a coupling constant that carries a
{\it negative} dimension of mass and this makes the operator
non-renormalizable. However, once the Chern-Simons interaction
turns the operator marginal or relevant, the effective coupling of
the four-fermion interaction has zero or a positive dimension and
therefore the four-fermion operator becomes renormalizable or
super-renormalizable.

It is equally interesting to conduct a parallel discussion
for the Chern-Simons boson model.
In this model, as the scalar field in three dimensions
has an engineering dimension $1/2$,
the boson self-interactions $g_1(\phi^*\phi)^2$ and
$g_2(\phi^*\phi)^3$ are relevant and marginal, respectively,
and therefore should be taken into account.
However, if one perturbs the theory near
the Gaussian (ultra-violet) fixed point of these interactions,
where $g_1 \sim g_2 \sim 0$,
the boson self-interactions can be turned off for the time being.
Assuming so, similar to (\ref{dn}), we obtain the scaling dimensions
for the operators $(\phi^*\phi)^n$
\begin{equation}
d_{(\phi^*\phi)^n} = n - {7n+6\over 6\kappa^2}
+ O({1\over \kappa^4}).
\label{dnb}
\end{equation}
Same to what happens in the fermion model,
the Chern-Simons interaction decreases
the scaling dimensions of these operators.
The critical Chern-Simons coupling
for a phase transition at which the leading
irrelevant operator $(\phi^*\phi)^4$ becomes marginal is
\begin{equation}
1/\kappa_c^2 = 3/17.
\label{kcb}
\end{equation}
On the other hand, if a perturbation is performed near the
infrared fixed points of the self-interactions $(\phi^*\phi)^2$
and/or $(\phi^*\phi)^3$, one can not switch off them,
as the critical couplings $g^*_1$ and $g^*_2$ can be rather strong
and they are the driving forces near the infrared fixed points.
The self-interactions contribute to (\ref{dnb}) {\it positive} terms that
increase the scaling dimensions of the operators.
The self-interaction of $(\phi^*\phi)^2$, for example,
at the infrared fixed point may be so strong
that it alone makes 
itself irrelevant.
However, a sufficiently strong Chern-Simons interaction
may possibly draw it back,
and the corresponding infrared fixed point is
significantly shifted.

Another interesting case is
the Chern-Simons matter system involving $N$ spices of matters
with some symmetry, $O(N)$ for instance. A possible perturbation
expansion in this case is over $1/N$.
We shall consider this expansion and see the same phenomenon at
the order $O(1/N)$, just before ending the letter.

Now we turn to renormalization.
Though, in a procedure of renormalization,
one normally deals with the ultra-violet
divergences, the resulting
(ultra-violet) finite effective theory takes a form
that is equally good in
exhibiting the asymptotic behavior of the theory
in both high and low energy limits.
Let us take a simple example. After renormalization, the two-point
function of fermion in the momentum space has an asymptotic form
%
$S(p) = <\psi(p)\bar\psi(-p)>
= \frac{1}{i\gamma\cdot p}(\frac{p^2}{\mu^2})^{\gamma_\psi},$
%
where $\mu$ is a reference mass parameter and $\gamma_\psi$
the anomalous dimension of the fermion field. In the Chern-Simons
fermion model,
$\gamma_\psi = -\frac{1}{3\kappa^2} \leq 0$,
at the lowest non-trivial order and in the Landau gauge\cite{CSW}.
This implies that the fermion field
in th Chern-Simons quantum field theory
has a dimension less than its engineering one.
Moreover, the kinetic term $\bar\psi\gamma\cdot\partial\psi$
takes an asymptotic form $(\frac{p^2}{\mu^2})^{\gamma_\psi}$. Now
we see the kinetic of fermion in the Chern-Simons theory
is relevant in the lower energy region, instead of marginal.

Though there are several independent terms in the action (\ref{action}),
to renormalize it, only {\it one} non-trivial renormalization
constant suffices. This is because
$Z_A = 1$ can always be chosen, as the Abelian gauge field $A$ needs no
infinite renormalization,
and $Z_\psi = Z_{(\bar\psi A\psi)}$, due to the gauge symmetry.
The fermion wave-function renormalization constant,
to the lowest non-trivial order, in the Landau gauge, and under the minimal
subtraction, is \cite{CSW}
\begin{equation}
Z_\psi=1+{ 1\over 3\kappa^2}{1\over \epsilon},
\label{z}
\end{equation}
with $\epsilon = 3 - \omega \rightarrow 0$. Now we consider the
composite operators of interest.
For simplicity, we shall set matter mass to zero ($m=0$).
This will not change the ultra-violet divergence
structure of the model. We shall use the Landau gauge for
convenience, the
results must be independent of a gauge choice as we are dealing
with the gauge
invariant operators. The regularization by dimensional
reduction,
as  used in
\cite{CSW} \cite{FCW}, will be used here. The same results have been obtained
by introducing a naive cutoff, as a check of consistency.
To calculate the renormalization of a composite operator
$O_n =\frac{1}{(n!)^2} (\bar\psi\psi)^n$ for a given $n$, we construct
an 1PI composite vertex $\Gamma_{O_n}^{2n}$ which contains the operator
$O_n$ as a vertex and has $n$ truncated external fermion and $n$ anti-fermion
(and null Chern-Simons) lines. The non-trivial
Feynman diagrams at the lowest non-trivial order are listed in Fig.~$1$.

\unitlength=1.00mm
\linethickness{0.4pt}

\begin{picture}(140.00,90.00)

\thicklines

\put(30.00,75.00){\circle*{4.00}}
\put(70.00,75.00){\circle*{4.00}}
\put(110.00,75.00){\circle*{4.00}}
\put(30.00,75.0){\line(-1,-2){14.00}}
\put(70.00,75.0){\line(-1,-2){14.00}}
\put(110.00,75.0){\line(-1,-2){14.00}}
\put(30.00,75.0){\line(1,-2){14.00}}
\put(70.00,75.0){\line(1,-2){14.00}}
\put(110.00,75.0){\line(1,-2){14.00}}
\multiput(23.50,61.50)(2.00,0.00){7}{\line(3,0){1.00}}
\multiput(18.50,51.50)(2.00,0.00){12}{\line(3,0){1.00}}
\multiput(63.50,61.50)(2.00,-1.00){9}{\line(3,0){1.00}}
\multiput(75.50,61.50)(-2.00,-1.00){9}{\line(3,0){1.00}}
\multiput(99.25,53.00)(2.00,0.00){3}{\line(3,0){1.00}}
\multiput(120.00,53.00)(-2.00,0.00){3}{\line(3,0){1.00}}
\put(110,53){\circle{10}}

\put(30.00,35.00){\circle*{4.00}}
\put(70.00,35.00){\circle*{4.00}}
\put(110.00,35.00){\circle*{4.00}}
\put(30.00,35.0){\line(-1,-2){14.00}}
\put(70.00,35.0){\line(-1,-2){14.00}}
\put(110.00,35.0){\line(-1,-2){7.00}}
\put(30.00,35.0){\line(1,-2){14.00}}
\put(70.00,35.0){\line(1,-2){14.00}}
\put(110.00,35.0){\line(1,-2){7.00}}
\multiput(33.0,28.5)(2.00,-1.00){4}{\line(3,0){1.0}}
\multiput(39.50,16.0)(0.00,2.00){5}{\line(3,0){1.00}}
\multiput(18.50,11.5)(2.00,0.00){12}{\line(3,0){1.00}}
\multiput(62.50, 19.5)(2.00,0.00){8}{\line(3,0){1.00}}
\multiput(75.0,25.5)(2.00,-1.00){4}{\line(3,0){1.0}}
\multiput(81.0,13.0)(0.00,2.00){5}{\line(3,0){1.00}}
\multiput(103.0,20.0)(0.00,-2.00){5}{\line(0,3){1.00}}
\multiput(117.0,20.0)(0.00,-2.00){5}{\line(0,3){1.00}}
\put(103.0,20.95){\line(1,0){14.}}
\put(103.0,12.00){\line(1,0){14.}}
\put(103.0,12.00){\line(-1,-2){2.7}}
\put(117.0,12.00){\line(1,-2){2.7}}
\end{picture}
\begin{description}
\item[Fig. 1]
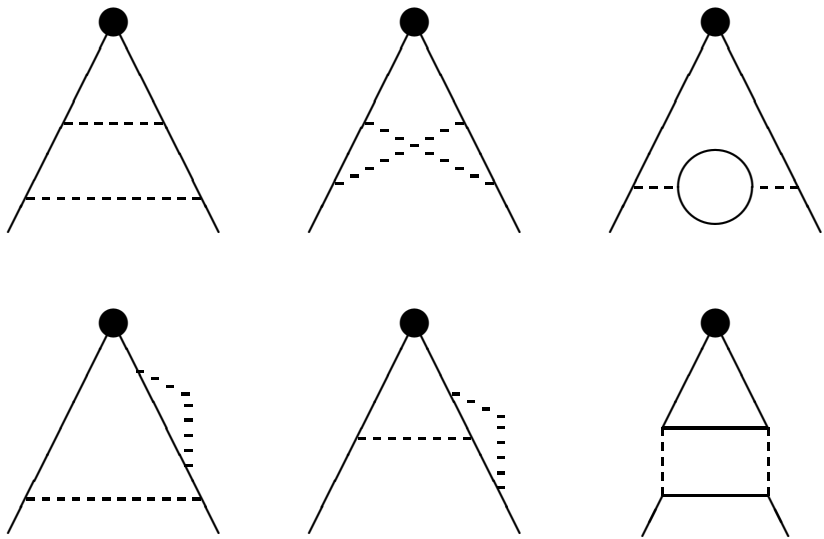
\ \ \ \ Non-trivial Feynman diagrams for $\Gamma^{2n}_{O_n}$
 at the order $O(1/\kappa^2)$ in the fermion model.
Real line stands for the fermion propagator;
dashed line the Chern-Simons propagator;
and dark spot the operator $O_n = \frac{1}{(n!)^2}(\bar\psi\psi)^n$
with $2(n-1)$ external fermion lines omitted.
A symmetric factor two for each of the last three diagrams.
\end{description}
By power counting, the composite vertex $\Gamma^{2n}_{O_n}$ is dimensionless.
The calculation is somehow tedious but straightforward. The result turns out
to be
\begin{equation}
\Gamma^{2n}_{O_n}({\bf 0},p;\frac{1}{\epsilon})
= 1 + {\bf [} \frac{5}{2\kappa^2}(\frac{1}{\epsilon})
+ ln(\frac{\mu^2}{p^2}) + finite{\bf ]},
\label{div}
\end{equation}
where, without loss of generality, we have set all external momenta but
one zero.  Then using the renormalization relation
\begin{equation}
(\Gamma^{2n}_{O_i})_R(p)
=Z_\psi^{n}(\frac{1}{\epsilon})
(Z_{O})_{ij}(\frac{1}{\epsilon})\Gamma^{2n}_{O_j}(p;\frac{1}{\epsilon})
\label{rela}
\end{equation}
and the fermion wave-function renormalization constant (\ref{z}),
we obtain the renormalization constant and anomalous dimension for
the operator $(\bar\psi\psi)^n$
\begin{eqnarray}
Z_{(\bar\psi\psi)^n} &=& 1 - \frac{2n+15}{6\kappa^2}(\frac{1}{\epsilon}),
\label{zon}\\
\gamma_{(\bar\psi\psi)^n} &=& - \frac{2n+15}{6\kappa^2}.\label{gon}
\end{eqnarray}
It is worth to notice a simplicity appeared here.
As shown in the renormalization relation (\ref{rela}), in its general form,
there may exist operator mixing in renormalization of higher dimensional
operators.
The operator mixing, if happens, will make the situation much more
complicated. However, operator mixing
does {\it not} appear in the renormalization of the class of
operators $(\bar\psi\psi)^n$ of the Chern-Simons fermion
model, as we have seen; neither in that of $(\phi^*\phi)^n$ of the Chern-Simons
boson model, as we shall see, at least in the lowest non-trivial order.
This is because the only primary divergence brought by the operator
$(\bar\psi\psi)^n$ for a given $n$ is that in (\ref{div}).
In other words, any cutoff dependence associated with the vertex $O_n$
can be cured by just one counterterm associated with (\ref{zon}). We have
calculated another composite vertex with $O_n$, $\Gamma^{2(n-1)}_{O_n}$,
and confirmed that there is indeed no need for any new counterterm to
cure the divergence - the one given in (\ref{zon}) suffices -  though the
composite vertex  $\Gamma^{2(n-1)}_{O_n}$ has dimension two and may have
various divergences.

In the Chern-Simons boson model, the gauge invariant composite operators under
consideration are $O_n = \frac{1}{(n!)^2}(\phi^*\phi)^n$, which have
an engineering dimension $n$.
To perform the renormalization and therefore to calculate their anomalous
dimensions, we consider similarly
the composite vertex $\Gamma^{2n}_{O_n}$ with the operator $O_n$
and $n$ external charged boson and $n$ anti-charged boson lines. The life here
seems easier as, at the lowest non-trivial order, there is only
one non-trivial diagram, as shown in Fig. $2$, for $\Gamma^{2n}_{O_n}$.
%
%
%

\unitlength=1.00mm
\linethickness{0.4pt}

\thicklines

\begin{picture}(160.00,50.00)
\put(70.00,35.00){\circle*{4.00}}
\put(70.00,35.0){\line(-1,-2){7.00}}
\put(70.00,35.0){\line(1,-2){7.00}}
\multiput(63.,20.0)(1.00,-2.00){8}{\line(0,3){1.00}}
\multiput(77.,20.0)(-1.00,-2.00){8}{\line(0,3){1.00}}
\put(63.0,21.){\line(1,0){14.}}
\put(63.0,6.){\line(1,0){14.0}}

\end{picture}
\begin{description}
\item[Fig. 2]
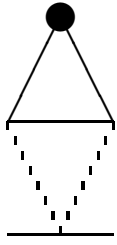
\ \ \ \ The only non-trivial Feynman diagram for $\Gamma^{2n}_{O_n}$
at the order $O(1/\kappa^2)$ in the boson model.
Real line stands for the boson propagator; dashed line the Chern-Simons
propagator; and dark spot the operator $O_n = \frac{1}{(n!)^2}
(\phi^*\phi)^n$ with $2(n-1)$
external boson lines omitted. There is a symmetric
factor two for the diagram.
\end{description}
Calculating Fig.~$2$, we obtain
\begin{equation}
\Gamma^{2n}_{O_n}({\bf 0},p;\frac{1}{\epsilon})
= 1 + {\bf [} \frac{1}{\kappa^2}(\frac{1}{\epsilon})
+ ln(\frac{\mu^2}{p^2}) + finite{\bf ]}.
\label{div1}
\end{equation}
Finally, we have the
renormalization constant and anomalous dimension of $(\phi^*\phi)^n$
\begin{eqnarray}
Z_{(\phi^*\phi)^n} &=& 1 -
\frac{7n+6}{6\kappa^2}(\frac{1}{\epsilon}),\label{zon1}\\
\gamma_{(\phi^*\phi)^n} &=& - \frac{7n+6}{6\kappa^2}.\label{gon1}
\end{eqnarray}
To get these, we have used the boson model version of the
renormalization relation
(\ref{rela}) and the boson wave-function renormalization constant
$Z_\phi = 1 + \frac{7}{6\kappa^2}(\frac{1}{\epsilon})$  \cite{CSW}.

To conclude this letter,  we discuss the $1/N$ expansion,
assuming
there are $N$ species of matter fields which obey a global symmetry
$O(N)$. We shall see the same phenomenon that the Chern-Simons interaction
makes
the composite operators of interest more relevant or less irrelevant. We
take the fermion model as an example and it is straightforward
to generalize to the boson model.
As the one-loop fermion bubble chain is at the same $O(1/N^0)$ order
with the bare Chern-Simons propagator, one must use a
dressed gauge propagator that sums over the one-fermion-loop chains,
instead of the bare one, in the $1/N$ expansion.
The dressed gauge propagator in the Landau gauge takes the
form, same in both the fermion and boson models,
\begin{eqnarray}
\Delta^{\mu\nu}(p) &=& A \frac{\delta^{\mu\nu}p^2-p^\mu p^\nu}{p^3}
+B\frac{\epsilon^{\mu\nu\lambda}p^\lambda}{p^2},\\
A&=& \frac{8\pi\theta}{64\theta^2 + \pi^2}, ~~~
B = - \frac{64\theta^2}{64\theta^2 + \pi^2},
\label{ab}
\end{eqnarray}
where $1/\theta$ is the effective Chern-Simons coupling.
The fermion wave-function renormalization constant at the order $O(1/N)$
and in the Landau gauge is \cite{FCW}
\begin{equation}
Z_{\psi_i}
=1 + \frac{16}{3\pi\theta(64\theta^2+\pi^2)N}(\frac{1}{\epsilon}).
\label{z1}
\end{equation}
The non-trivial diagrams, at the same $O(1/N)$ order, for the composite vertex
$\Gamma^{2n}_{O_n}$ with the gauge invariant composite operator
$\frac{1}{(n!)^2}(\bar\psi_i\psi_i)^n$
are given in Fig.~$3$.
\unitlength=1.00mm
\linethickness{0.4pt}

\begin{picture}(160.00,50.00)
\thicklines
\put(30.00,35.00){\circle*{4.00}}
\put(30.00,35.0){\line(-1,-2){13.50}}
\put(30.00,35.0){\line(1,-2){13.50}}
\multiput(21.50,17.50)(2.00,0.00){9}{\line(3,0){1.00}}
\multiput(20.50,16.50)(2.00,0.00){10}{\line(3,0){1.00}}
\put(90.00,35.00){\circle*{4.00}}
\put(90.00,35.0){\line(-1,-2){7.00}}
\put(90.00,35.0){\line(1,-2){7.00}}
\multiput(83.0,20.0)(0.00,-2.00){4}{\line(0,3){1.00}}
\multiput(97.0,20.0)(0.00,-2.00){4}{\line(0,3){1.00}}
\multiput(84.0,20.0)(0.00,-2.00){4}{\line(0,3){1.00}}
\multiput(96.0,20.0)(0.00,-2.00){4}{\line(0,3){1.00}}
\put(83.0,20.95){\line(1,0){14.}}
\put(83.0,14.00){\line(1,0){14.}}
\put(83.0,14.00){\line(-1,-2){3.0}}
\put(97.0,14.00){\line(1,-2){3.0}}
%
%
\put(30, 0.0){\makebox(0,)[cc]{$(a)$}}
\put(90, 0.0){\makebox(0,)[cc]{$(b)$}}

\end{picture}

\begin{description}
\item[Fig. 3]
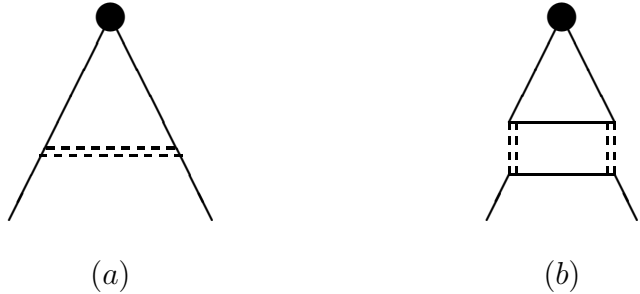
\ \ \ \ Non-trivial Feynman diagram of $\Gamma^{2n}_{O_n}$
at the order $O(1/N)$.
Real line stands for the fermion propagator; double dashed line the dressed
gauge propagator; and dark spot the operator
$O_n = \frac{1}{(n!)^2}(\bar\psi_i\psi_i)^n$ with $2(n-1)$
external lines omitted.
There is a symmetric factor two for the diagram $(b)$.
\end{description}
Calculating the diagrams in Fig.~$3$, and using (\ref{rela}), (\ref{ab}) and
(\ref{z1}), we obtain at the order $O(1/N)$
\begin{equation}
\gamma_{(\bar\psi\psi)^n}
= - \frac{16(64n\theta^2+n\pi^2+96\theta^2-9\pi^2)}{3(64\theta^2+\pi^2)^2}
\frac{1}{N}.
\end{equation}
Now we see the anomalous dimensions of $(\bar\psi\psi)^n$ are negative
and therefore these operators are more relevant (for $n=1$) or less
irrelevant (for $n\geq 2$), only if the Chern-Simons coupling is not
unreasonably strong, not stronger than $1/\theta^2 \sim 15$ for $n=2$ for
instance.

The authors would thank I. Affleck, J.-W. Gan, G. Semenoff, Y.-S. Wu
for discussions. Work of W.C. was supported in part by the Natural Sciences
and Engineering Research Council of Canada. Work of M.L. was supported
by DOE grant DOE-76ER70023. Part of this work was done during M.L. visiting
Aspen Center for Physics, its hospitality is gratefully acknowledged.

\end{document}